\documentclass[10pt,letterpaper,revtex]{article}
\usepackage[dvips]{graphics}

\makeatletter

\newcommand{\LyX}{L\kern-.1667em\lower.25em\hbox{Y}\kern-.125emX\spacefactor1000}

\newcommand{\lyxaddress}[1]{
  \par {\raggedright #1 
  \vspace{1.4em}
  \noindent\par}
}

\makeatother

\begin{document}

\title{Stokes parameters for light scattering from a Faraday-active sphere}

\author{D. Lacoste and B. A. van Tiggelen}

\maketitle

\lyxaddress{\emph{Laboratoire de Physique et Modélisation des Milieux Condensés, Maison
des Magistères, B.P. 166, 38042 Grenoble Cedex 9, France}}

\begin{abstract}
We \emph{}present an exact calculation for the scattering of light from a single
sphere made of Faraday-active material, to first order in the external magnetic
field. We use a recent expression for the T-matrix of a Mie scatterer in a magnetic
field to compute the Stokes parameters in single scattering that describe flux
and polarization of the scattered light.
\end{abstract}

\section{Introduction}

Several reasons exist why one wishes to understand light scattering from a dielectric
sphere made of magneto-active material. Single scattering is the building block
for multiple scattering. Recently, many experiments such as those reported by
Erbacher \emph{etal.} \cite{maret} and Rikken \emph{etal.} \cite{rikken} have
been done with diffuse light in a magnetic field. It turns out that the theory
using point-like scatterers in a magnetic field, as first developed by MacKintosh
and John \cite{john}, does not always enable a quantitative analysis, for the
evident reason that experiments do not contain ``small'' scatterers. This
paper addresses light scattering from a sphere of any size in a homogeneous
magnetic field.

The model of Rayleigh scatterers was used successfully to describe specific
properties of multiple light scattering in magnetic fields, such as for instance
Coherent Backscattering, \emph{Photonic Hall Effect (PHE)} and \emph{Photonic
Magneto-resistance (PMR)}. The first study of one gyrotropic sphere, due to
Ford and Werner \cite{ford}, was applied to the scattering of semiconducting
spheres by Dixon and Furdyna \cite{dixon}. For the case of magneto-active particles,
for which the change in the dielectric constant induced by the magnetic field
is small, a perturbational approach is in fact sufficient. Kuzmin \cite{kuzmin}
showed that the problem of scattering by a weakly anisotropic particle of any
type of anisotropy can be solved to first order in the perturbation. Using a
T-matrix formalism, Lacoste \emph{etal.} \cite{josa} developed independantly
a perturbational approach for the specific case of magneto-optical anisotropy.
This was successfully applied to compute the diffusion coefficient for magneto-transverse
light diffusion \cite{euro}. Using the T-matrix for a Mie scatterer in a magnetic
field we have obtained, we discuss the consequences for the Stokes parameters
\cite{hulst} that describe the polarization of the scattered light.

\section{Perturbation theory}

In this paper we set \( c_{0}=1 \). In a magnetic field \( \mathbf{B} \),
the refractive index is a tensor of rank two. For the standard Mie problem,
its value at position \( \mathbf{r} \) depends on the distance to the center
of the sphere \( |\mathbf{r}| \), which has a radius \( a \) via the Heaviside
function \( \Theta (|\mathbf{r}|-a), \) that equals 1 inside the sphere and
0 outside, \textbf{
\begin{eqnarray}
{\bf \varepsilon }(\mathbf{B},\mathbf{r})-\mathbf{I}= & \left[ (\varepsilon _{0}-1)\, \mathbf{I}+\varepsilon _{F}\, \, {\bf \Phi }\right] \Theta (|\mathbf{r}|-a).\label{epsilon} 
\end{eqnarray}
}In this expression, \( \mathbf{I} \) is the identity tensor, \( \varepsilon _{0}=m^{2} \)
is the value of the normal isotropic dielectric constant of the sphere of relative
index of refraction \( m \) (which is allowed to be complex-valued) and \( \varepsilon _{F}=2mV_{0}B/\omega  \)
is a dimensionless coupling parameter associated with the amplitude of the Faraday
effect (\( V_{0} \) being the Verdet constant, \( B \) the amplitude of the
magnetic field, and \( \omega  \) the frequency). We introduced the antisymmetric
hermitian tensor \( \Phi _{ij}=i\epsilon _{ijk}\hat{B}_{k} \) (the hat above
vectors notes normalized vectors). The Mie solution depends on the dimensionless
size parameters \( x=\omega a \) and \( y=mx \). In this paper we restrict
ourselves to non-absorbing media so that \( m \) and \( \varepsilon _{F} \)
are real-valued. Since \( \varepsilon _{F}\approx 10^{-4} \) in most experiments,
a perturbational approach is valid.

Upon noting that the Helmholtz equation is formally analogous to a Schrödinger
equation with potential \( \mathbf{V}(\mathbf{r},\omega )=\left[ \mathbf{I}-\varepsilon (\mathbf{B},\mathbf{r})\right] \, \omega ^{2} \)
and energy \( \omega ^{2} \), the T-operator is given by the following Born
series,

\begin{equation}
\label{Born}
\mathbf{T}(\mathbf{B},\mathbf{r},\omega )=\mathbf{V}(\mathbf{r},\omega )+\mathbf{V}(\mathbf{r},\omega )\cdot \mathbf{G}_{0}(\omega ,\mathbf{p})\cdot \mathbf{V}(\mathbf{r},\omega )+...
\end{equation}
Here \( \mathbf{G}_{0}(\omega ,\mathbf{p})=1/(\omega ^{2}\mathbf{I}-p^{2}{\bf \Delta _{p}}) \)
is the free Helmholtz Green's operator in gaussian rationalized units for pure
dielectric particles, and \( \mathbf{p}=-i\nabla  \) is the momentum operator.
The tensor of rank two \( (\Delta _{p})_{ij}=\delta _{ij}-p_{i}p_{j}/p^{2} \)
projects upon the space transverse to the direction of \( \mathbf{p} \). The
T-matrix is defined as,

\begin{equation}
\label{tmatrix}
\mathbf{T}_{\mathbf{k}\sigma ,\mathbf{k}'\sigma '}=<\mathbf{k},\sigma \, |\, \mathbf{T}\, |\, \mathbf{k}',\sigma '>,
\end{equation}
where \( |\, \mathbf{k},\sigma > \)(respectively \( |\, \mathbf{k}',\sigma '> \))
represents an incident (respectively emergent) plane wave with direction \( \mathbf{k} \)
and state of helicity \( \sigma  \) (respectively \( \mathbf{k}' \) and \( \sigma ' \)).We
will call \( \mathbf{T}^{0} \) the part of \( \mathbf{T} \) that is independent
of the magnetic field and \( \mathbf{T}^{1} \) the part of the T-matrix linear
in \( \mathbf{B} \). We have found the following result \cite{josa},

\begin{equation}
\label{psi}
\mathbf{T}_{\mathbf{k}\sigma ,\mathbf{k}'\sigma '}^{1}=\varepsilon _{F}\, \omega ^{2}<\Psi ^{-}_{\mathbf{k},\sigma }\, |\, \Theta \, {\bf \Phi }\, |\, \Psi _{\mathbf{k}',\sigma '}^{+}>,
\end{equation}
 where the \( \Psi _{\mathbf{k},\sigma }^{\pm }\, (\bf r) \) are the unperturbed
eigenfunctions of the conventional Mie problem. This eigenfunction represents
the electric field at the point \( \bf r \) for an incident plane wave \( |\, \mathbf{k},\sigma > \).
This eigenfunction is ``outgoing'' for \( \Psi _{\mathbf{k},\sigma }^{+} \)
and ``ingoing'' for \( \Psi _{\mathbf{k},\sigma }^{-} \). Eq. (\ref{psi})
resembles the perturbation formula for the Zeeman shift in terms of the atomic
eigenfunctions, although here it provides a complex-valued amplitude in terms
of \emph{continuum} eigenfunctions, rather than a real-valued energy shift in
terms of bound states.

\section{T matrix for Mie scattering}

In order to separate radial and an angular contribution in Eq. (\ref{psi}),
we used a well-known expansion of the Mie eigenfunction \( \Psi _{\mathbf{k},\sigma }^{+} \)
in the basis of vector spherical harmonics \cite{newton}. We choose the quantification
axis \( z \) along the magnetic field. With this choice, the operator \( \mathbf{S}_{z} \),
the \( z \)-component of a spin one operator, can be associated with the tensor
\( {\bf -\Phi } \). The eigenfunctions of the operator \( \mathbf{S}_{z} \)
form a convenient basis for the problem. The expansion of Eq. (\ref{psi}) in
vector spherical harmonics leads to a summation over quantum numbers \( J,J',M \)
and \( M' \). The Wigner-Eckart theorem applied to the vector operator \( \mathbf{S} \)
gives the selection rules for this case \( J=J' \) and \( M=M' \).

The radial integration can be done using a method developed by Bott \emph{etal}.
\cite{bott}, which gives,

\begin{equation}
\label{T1}
\mathbf{T}_{\mathbf{k},\mathbf{k}'}^{1}=\frac{16\pi }{\omega }\, W\, \sum _{J,M}\, (-M)\, \left[ \mathcal{C}_{J}\, \mathbf{Y}_{J,M}^{e}(\hat{\mathbf{k}})\, \mathbf{Y}^{e*}_{J,M}(\hat{\mathbf{k}}')+\mathcal{D}_{J}\, \mathbf{Y}^{m}_{J,M}(\hat{\mathbf{k}})\, \mathbf{Y}^{m*}_{J,M}(\hat{\mathbf{k}}')\right] ,
\end{equation}
with the dimensionless parameter:

\[
W=V_{0}B\lambda ,\]
\( \lambda  \) is the wavelength in the medium. The meaning of the indices
\( e,m \) is explained in the Appendix. In the limiting case of a perfect dielectric
sphere with no absorption ( \( \Im m(m)\rightarrow 0 \) ), the coefficients
are given by,

\begin{equation}
\label{C2}
\mathcal{C}_{J}=\frac{-c^{2*}_{J}u_{J}^{2}\, y}{J(J+1)}\, \left( \frac{A_{J}}{y}-\frac{J(J+1)}{y^{2}}+1+A^{2}_{J}\right) ,
\end{equation}

\begin{equation}
\label{D2}
\mathcal{D}_{J}=\frac{-d^{2*}_{J}u_{J}^{2}\, y}{J(J+1)}\, \left( -\frac{A_{J}}{y}-\frac{J(J+1)}{y^{2}}+1+A^{2}_{J}\right) ,
\end{equation}
 with \( A_{J}\, (y)=u_{J}'\, (y)/u_{J}\, (y) \), \( u_{J}(y) \) the Ricatti-Bessel
function, and \( c_{J} \) and \( d_{J} \) the Mie amplitude coefficients of
the internal field \cite{hulst}.

Two important symmetry relations must be obeyed by our T-matrix. The first one
is parity symmetry and the second one reciprocity. These relations can be established
generally when the hamiltonian of a given system has the required symmetries
( \emph{cf.} Eq. (15.53) and Eq.(15.59a) P454 of Ref. (\cite{newton}) ). We
give in the Appendix a less general derivation of these relations for our specific
problem:

\begin{equation}
\label{parity}
T_{-\mathbf{k}\sigma ,-\mathbf{k}'\sigma '}(\mathbf{B})=T_{\mathbf{k}\sigma ,\mathbf{k}'\sigma '}(\mathbf{B}),
\end{equation}

\begin{equation}
\label{reciprocity}
T_{-\mathbf{k}'-\sigma ',-\mathbf{k}-\sigma }(-\mathbf{B})=T_{\mathbf{k}\sigma ,\mathbf{k}'\sigma '}(\mathbf{B}).
\end{equation}
 We emphasize that \( \sigma (-\mathbf{k})=-\sigma (\mathbf{k}) \), i.e. \( \sigma  \)
indicates in fact helicity and \emph{not} circular polarization. The helicity
is the eigenvalue of the operator \( \mathbf{S}\cdot \mathbf{k} \).

\subsection{The amplitude matrix }

The amplitude matrix \( \mathbf{A} \) relates incident and scattered field
with respect to an arbitrary plane of reference. A common choice is the plane
that contains the incident and the scattered wave vector, and which is for this
reason called the scattering plane. We will call the linear base the base made
of one vector in this plane and one vector perpendicular to it. In this basis,
the amplitude matrix sufficiently far \( \omega r\gg 1 \), is simply defined
from the T-matrix by,

\begin{equation}
\label{amatrix}
\mathbf{A}_{\mathbf{k},\mathbf{k}'}=\frac{-1}{4\pi r}\mathbf{T}^{*}_{\mathbf{k},\mathbf{k}'}=\frac{e^{i\phi }}{i\omega r}\left( \begin{array}{cc}
S_{2} & S_{3}\\
S_{4} & S_{1}
\end{array}\right) .
\end{equation}
\( \phi  \) is a phase factor that depends on the relative phase of the scattered
wave with respect to the incident wave and which is defined in Ref. (\cite{hulst}).
The complex conjugation in Eq. (\ref{amatrix}) is simply due to a different
sign convention in Newton \cite{newton}. When no magnetic field is applied,
the T-matrix of the conventional Mie-problem is given by a formula analogous
to Eq. (\ref{T1}) where \( \mathcal{C}_{J} \) and \( \mathcal{D}_{J} \) have
been replaced by the Mie coefficients \( a_{J} \) and \( b_{J} \), and with
\( M=1 \). Because of rotational invariance of the scatterer, the final result
only depends on \( \cos \theta  \), the scalar product of \( \mathbf{k} \)
and \( \mathbf{k}' \), \( \theta  \) is the scattering angle. Therefore we
get in the circular basis (associated with the helicities \( \sigma  \) and
\( \sigma ' \)):

\begin{equation}
\label{t0}
T_{\sigma \sigma '}^{0}=\frac{2\pi }{i\omega }\sum _{J\geq 1}\frac{2J+1}{J(J+1)}\, (a^{*}_{J}+\sigma \sigma 'b^{*}_{J})\, \left[ \pi _{J,1}(\cos \theta )+\sigma \sigma '\tau _{J,1}(\cos \theta )\right] .
\end{equation}
Alternatively the T-matrix may be expanded on the basis of the Pauli matrices

\begin{equation}
\label{pauli}
\mathbf{T}^{0}=\frac{2\pi }{i\omega }\left[ (S^{*}_{1}+S^{*}_{2})\mathbf{I}+(S^{*}_{1}-S^{*}_{2})\sigma _{x}\right] .
\end{equation}
In Eq. (\ref{t0}), the polynomials \( \pi _{J,M} \) and \( \tau _{J,M} \)
are defined in terms of the Legendre polynomials \( P^{M}_{J} \) by \cite{hulst},

\begin{equation}
\label{poly}
\pi _{J,M}(\theta )=\frac{M}{\sin \theta }P^{M}_{J}(\cos \theta )\, ,\, \, \, \, \, \, \, \, \, \, \, \, \, \, \, \, \, \tau _{J,M}(\theta )=\frac{d}{d\theta }P^{M}_{J}(\cos \theta ).
\end{equation}
 For \( M=1 \), \( \pi _{J,1} \) and \( \tau _{J,1} \) are polynomials of
\( \cos \theta  \) of order \( J-1 \) and \( J \) respectively but not in
general for any value of \( M \). When written in the linear basis of polarization,
Eq. (\ref{t0}) implies that a Mie scatterer has \( S_{3}=S_{4}=0 \) as imposed
by the rotational symmetry. For the backward direction \( \theta =\pi  \),
the reciprocity symmetry implies that \( S_{3}+S_{4}=0 \) for an arbitrary
particle (possibly non-spherical) \cite{hulst}. We will see that these two
properties do not hold anymore when a magnetic field is present.

\subsection{General case for \protect\( \mathbf{T}^{1}\protect \) when \protect\( \hat{\mathbf{k}}\times \hat{\mathbf{k}}'\neq 0\protect \) }

\begin{figure}
{\centering \rotatebox{-90}{\includegraphics{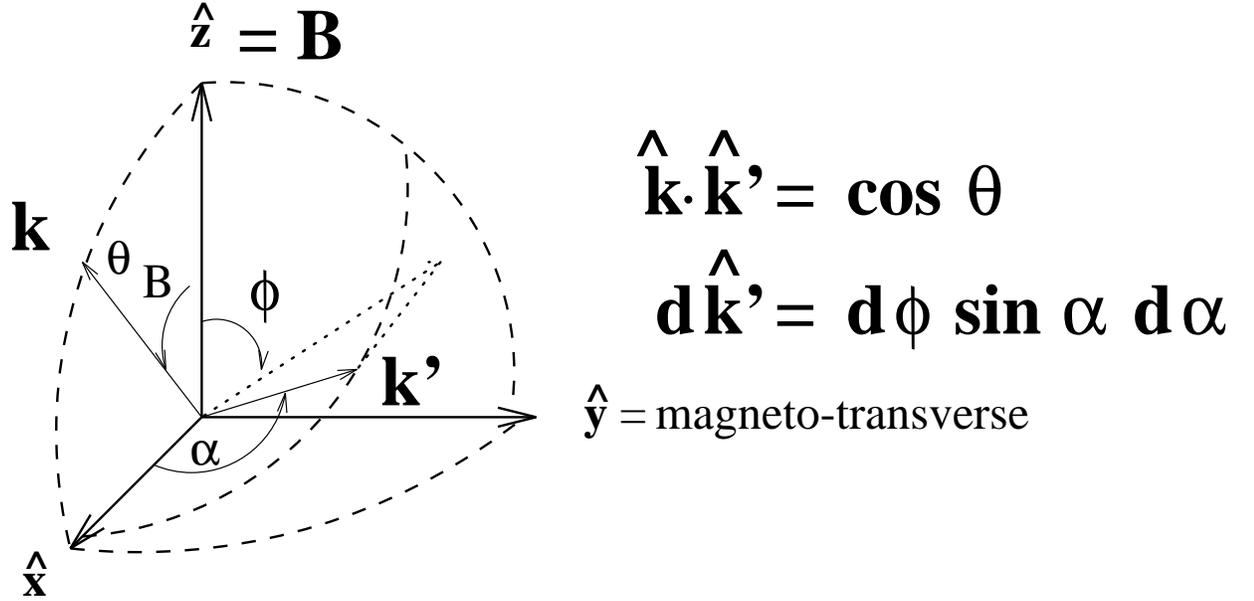}} \par}

\caption{Schematic view of the magneto-scattering geometry. Generally, \protect\( \theta \protect \)
denotes the angle between incident and outgoing wave vectors, \protect\( \phi \protect \)
is the azimuthal angle in the plane of the magnetic field and the \protect\( y\protect \)-axis.
This latter is by construction the magneto-transverse direction defined as the
direction perpendicular to both the magnetic field and the incident wave vector.
Angle \protect\( \alpha \protect \) coincides with angle \protect\( \theta \protect \)
in the special but relevant case that the incident vector is normal to the magnetic
field.\label{scheme}}
\end{figure}

It remains to express the vector spherical harmonics in Eq. (\ref{T1}), as
a function of the natural angles of the problem. In Fig. \ref{scheme}, we give
a schematic view of the geometry. In the presence of a magnetic field, the rotational
invariance is broken because \( \mathbf{B} \) is fixed in space. Because our
theory treats \textbf{\( \mathbf{T}^{1} \)} linear in \( \hat{\mathbf{B}} \),
\( \mathbf{T}^{1} \) can be constructed by considering only three special cases
for the direction of \( \hat{\mathbf{B}} \). If \( \hat{\mathbf{k}} \) and
\( \hat{\mathbf{k}}' \) are not collinear, we can decompose the unit vector
\( \hat{\mathbf{B}} \) in the non-orthogonal but complete basis of \( \hat{\mathbf{k}},\, \hat{\mathbf{k}}' \)
and \( \hat{\mathbf{g}}=\hat{\mathbf{k}}\times \hat{\mathbf{k}}'/|\hat{\mathbf{k}}\times \hat{\mathbf{k}}'| \).
This results in,

\begin{eqnarray}
\mathbf{T}_{\mathbf{kk}'}^{1} & = & \frac{(\hat{\mathbf{B}}\cdot \hat{\mathbf{k}})(\hat{\mathbf{k}}\cdot \hat{\mathbf{k}}')-\hat{\mathbf{B}}\cdot \hat{\mathbf{k}}'}{(\hat{\mathbf{k}}\cdot \hat{\mathbf{k}}')^{2}-1}\mathbf{T}_{\hat{\mathbf{B}}=\hat{\mathbf{k}}'}^{1}\nonumber \label{decomp} \\
 & + & \frac{(\hat{\mathbf{B}}\cdot \hat{\mathbf{k}}')(\hat{\mathbf{k}}\cdot \hat{\mathbf{k}}')-\hat{\mathbf{B}}\cdot \hat{\mathbf{k}}}{(\hat{\mathbf{k}}\cdot \hat{\mathbf{k}}')^{2}-1}\mathbf{T}_{\hat{\mathbf{B}}=\hat{\mathbf{k}}}^{1}\nonumber \label{decomp1} \\
 & + & (\hat{\mathbf{B}}\cdot \hat{\mathbf{g}})\mathbf{T}_{\hat{\mathbf{B}}=\hat{\mathbf{g}}}^{1},\label{Tdef} 
\end{eqnarray}
 The cases where \( \hat{\mathbf{B}} \) is either along \( \hat{\mathbf{k}} \)
or \( \hat{\mathbf{k}}' \) turn out to take the form,

\begin{equation}
\label{tbpk}
T_{\sigma \sigma '}^{1}(\hat{\mathbf{B}}=\hat{\mathbf{k}})=\frac{\pi }{\omega }[R_{1}(\cos \theta )\sigma +R_{2}(\cos \theta )\sigma '],
\end{equation}

\begin{equation}
\label{tbpkp}
T_{\sigma \sigma '}^{1}(\hat{\mathbf{B}}=\hat{\mathbf{k}}')=\frac{\pi }{\omega }[R_{1}(\cos \theta )\sigma '+R_{2}(\cos \theta )\sigma ],
\end{equation}

with

\begin{equation}
\label{p1}
R_{1}(\cos \theta )=-\frac{2W}{\pi }\sum _{J\geq 1}\frac{2J+1}{J(J+1)}\, [\mathcal{C}_{J}\pi _{J,1}(\cos \theta )+\mathcal{D}_{J}\tau _{J,1}(\cos \theta )]
\end{equation}

\begin{equation}
\label{p2}
R_{2}(\cos \theta )=-\frac{2W}{\pi }\sum _{J\geq 1}\frac{2J+1}{J(J+1)}\, [\mathcal{D}_{J}\pi _{J,1}(\cos \theta )+\mathcal{C}_{J}\tau _{J,1}(\cos \theta )]
\end{equation}
 In Ref. \cite{josa} we gave an expression for \( \mathbf{T}_{\sigma \sigma '}^{1}(\hat{\mathbf{B}}=\hat{\mathbf{g}}) \)
involving a double summation over the partial wave number \( J \) and the magnetic
quantum number \( M \). It is actually possible to do the summation over \( M \)
explicitely, thus simplifying considerably the numerical evaluation. Indeed,
if one expresses \( \mathbf{T}^{0} \) with respect to a \( z \)-axis perpendicular
to the scattering plane for a given partial wave \( J \), one ends up with
the following relation between the polynomials \( \pi _{J,M} \) and \( \tau _{J,M} \),

\begin{equation}
\label{pirel}
\pi _{J,1}(\cos \theta )=2\sum _{J\geq M\geq 1}\left[ \frac{(J-M)!}{(J+M)!}\tau _{J,M}(0)^{2}\cos (M\theta )\right] +\tau _{J,0}(0)^{2}
\end{equation}

\begin{equation}
\label{pirel2}
\tau _{J,1}(\cos \theta )=2\sum _{J\geq M\geq 1}\left[ \frac{(J-M)!}{(J+M)!}\pi _{J,M}(0)^{2}\cos (M\theta )\right] +\pi _{J,0}(0)^{2}
\end{equation}
 Upon performing the derivatives of these relations with respect to \( \theta  \)
and comparing to the expression for \( \mathbf{T}_{\sigma \sigma '}^{1}(\hat{\mathbf{B}}=\hat{\mathbf{g}}) \)
we find,

\begin{equation}
\label{tbg}
T_{\sigma \sigma '}^{1}(\hat{\mathbf{B}}=\hat{\mathbf{g}})=\frac{\pi }{\omega }(Q_{1}(\theta )+\sigma \sigma 'Q_{2}(\theta ))
\end{equation}
 with

\begin{equation}
\label{q1}
Q_{l}(\theta )=-i\frac{d}{d\theta }R_{l}(\cos \theta )=i\sin \theta \frac{d}{d\cos \theta }R_{l}(\cos \theta ),\, \, \, \, \, \, \, \, \, \, \, \, \, \, \, \, \, l=1,2.
\end{equation}
 We are convinced that a rigorous group symmetry argument exists that relates
the derivative of \( T_{\sigma \sigma '}^{1}(\hat{\mathbf{B}}=\hat{\mathbf{k}}) \)
with respect to \( \theta  \) to \( T_{\sigma \sigma '}^{1}(\hat{\mathbf{B}}=\hat{\mathbf{g}}). \)

\subsection{Particular case for \protect\( \mathbf{T}^{1}\protect \) when \protect\( \hat{\mathbf{k}}=\hat{\mathbf{k}}'\protect \)
and \protect\( \hat{\mathbf{k}}=-\hat{\mathbf{k}}'\protect \)}

The treatment in section (3.2) becomes degenerate when \( \hat{\mathbf{k}} \)
and \( \hat{\mathbf{k}}' \) are collinear, \textit{i.e.} in forward or backward
direction. In these cases, \( \hat{\mathbf{B}} \) can still be expressed on
a basis made of \( \hat{\mathbf{k}} \) and of two vectors perpendicular to
\( \hat{\mathbf{k}} \). The contribution of these last two vectors has the
same form as in \( \mathbf{T}_{\sigma \sigma '}^{1}(\hat{\mathbf{B}}=\hat{\mathbf{g}}) \)
for \( \theta =0 \) or \( \theta =\pi  \), which vanishes. An alternative
derivation consists to take the limit \( \theta \rightarrow 0 \) so that \( R_{1}=R_{2} \)
or \( \theta \rightarrow \pi  \) so that \( R_{1}=-R_{2} \) in Eqs. (\ref{tbpk}-\ref{p2}).
This yields,

\begin{equation}
\label{p10}
R_{1}(1)=R_{2}(1)=-\frac{W}{\pi }\sum _{J\geq 1}(2J+1)\, (\mathcal{C}_{J}+\mathcal{D}_{J})
\end{equation}
and 
\begin{equation}
\label{p1pi}
R_{1}(-1)=-R_{2}(-1)=-\frac{W}{\pi }\sum _{J\geq 1}(-1)^{J+1}(2J+1)\, (\mathcal{C}_{J}-\mathcal{D}_{J})
\end{equation}
This means,

\begin{equation}
\label{tkk}
\mathbf{T}_{\mathbf{k},\mathbf{k}}^{1}={\bf \Phi }\, \frac{2\pi }{\omega }R_{1}(1),
\end{equation}
and

\begin{equation}
\label{t-k-k}
\mathbf{T}_{\mathbf{k},-\mathbf{k}}^{1}={\bf \Phi }\, \frac{2\pi }{\omega }R_{1}(-1).
\end{equation}
 Both T-matrices contain the tensor \( {\bf \Phi } \) introduced in Eq. (\ref{epsilon})
for the dielectric constant of the medium of the sphere. For these two cases,
an operator can be associated with these T-matrices, which is \( \mathbf{S}_{\mathbf{z}} \)
since we have chosen \( \mathbf{B} \) along the \( z \)-axis. For \( \mathbf{T}_{\mathbf{k},\mathbf{k}}^{1} \),
the presence of the tensor \( {\bf \Phi } \) is to be expected since we know
that the forward scattering amplitude can be interpreted as an effective refractive
index in a transmission experiment \cite{hulst}. In the framework of an effective
medium theory, the real part of Eq. (\ref{tkk}) gives the Faraday effect whereas
the imaginary part gives the magneto-dichroism (\emph{i.e} different absorption
for different circular polarization) of an ensemble of Faraday-active scatterers.

\section{Magneto-transverse Scattering}

From \( \mathbf{T}^{1} \) matrix, we can compute how the magnetic field affects
the differential scattering cross section (summed over polarization) as a function
of the scattering angle. Its form can be guessed before doing any calculation
at all, since it must satisfy mirror-symmetry and the reciprocity relation \( d\sigma /d\Omega (\mathbf{k}\rightarrow \mathbf{k}',\mathbf{B})=d\sigma /d\Omega (-\mathbf{k}'\rightarrow -\mathbf{k},-\mathbf{B}) \).
A magneto-cross-section proportional to \( \hat{\mathbf{B}}\cdot \hat{\mathbf{k}} \)
or to \( \hat{\mathbf{B}}\cdot \hat{\mathbf{k}}' \) is parity forbidden since
\( \mathbf{B} \) is a pseudo-vector. Together with the rotational symmetry
of the sphere the only possibility is:

\begin{equation}
\label{sectioneff}
\frac{d\sigma }{d\Omega }(\mathbf{k}\rightarrow \mathbf{k}',\mathbf{B})=F_{0}(\cos \theta )+\det (\hat{\mathbf{B}},\hat{\mathbf{k}},\hat{\mathbf{k}}')F_{1}(\cos \theta )
\end{equation}
where \( \det (\mathbf{A},\mathbf{B},\mathbf{C})=\mathbf{A}\cdot (\mathbf{B}\times \mathbf{C}) \)
is the scalar determinant constructed from these three vectors. The second term
in Eq. (\ref{sectioneff}) will be called the magneto-cross-section. 

The magneto-cross section implies that there may be more photons scattered ``upwards''
than ``downwards'', both directions being defined with respect to the magneto-transverse
vector \( \hat{\mathbf{k}}\times \hat{\mathbf{B}} \) perpendicular to both
incident wave vector and magnetic field. An easy calculation yields,

\begin{equation}
\label{phe}
\Delta \sigma =\sigma _{up}-\sigma _{down}=\pi \int ^{\pi }_{0}d\theta \sin ^{3}\theta F_{1}(\cos \theta )
\end{equation}
A non-zero value for \( \Delta \sigma  \) will be referred to as a \emph{Photon
Hall Effect} (PHE). 

For Rayleigh scatterers, the above theory simplifies dramatically because one
only needs to consider the first partial wave of \( J=1 \) and the first terms
in a development in powers of \( y \) ( since \( y\ll 1 \) ). From Eqs. (\ref{C2})
and (\ref{D2}) we find that, \( \mathcal{C}_{1}=-2y^{3}/m^{2}(2+m^{2})^{2} \)
and \( \mathcal{D}_{1}=-y^{5}/45m^{4} \). We can keep only \( \mathcal{C}_{1} \)
and drop \( \mathcal{D}_{1} \) as a first approximation. Adding all the contributions
of Eqs. (\ref{decomp}) and (\ref{t0}), we find, in the linear base

\begin{equation}
\label{ttotal}
\mathbf{T}_{\mathbf{k},\mathbf{k}'}=\left( \begin{array}{cc}
t_{0}\hat{\mathbf{k}}\cdot \hat{\mathbf{k}}'+it_{1}\hat{\mathbf{B}}\cdot (\hat{\mathbf{k}}\times \hat{\mathbf{k}}') & -it_{1}\hat{\mathbf{B}}\cdot \hat{\mathbf{k}}\\
it_{1}\hat{\mathbf{B}}\cdot \hat{\mathbf{k}}' & t_{0}
\end{array}\right) .
\end{equation}
 where \( t_{0}=-6i\pi a^{*}_{1}/\omega  \) and \( t_{1}=-6C_{1}W/\omega  \).
This form agrees with the Rayleigh point-like scatterer discussed in Ref. \cite{bart}. 

A magnetic field breaks the rotational symmetry of the particle. If it is contained
in the scattering plane, Eq. (\ref{ttotal}) shows that we must have a non-zero
value for \( S_{3} \) and \( S_{4} \) as opposed to the case when no magnetic
field is applied. This property still holds for a Mie scatterer, the difference
being only present in the angular dependence of the elements of the amplitude
matrix. A magnetic field also violates the standard reciprocity principle as
can be seen on Eq. (\ref{reciprocity}). This implies that \( S_{3}+S_{4} \)
is non zero for the backward direction \( \theta =\pi  \). The relation \( S_{3}+S_{4}=0 \)
for the backward direction was derived by Van Hulst, but does not apply when
a magnetic field is present. In fact the magnetic field imposes that \( S_{3}=S_{4} \)
at backscattering. This is readily confirmed by the Rayleigh particle, for which
Eq. (\ref{ttotal}) implies that \( S_{3}+S_{4}=2S_{3}=-2it_{1}\hat{\mathbf{B}}\cdot \hat{\mathbf{k}} \)
for \( \theta =\pi  \). 

\begin{figure}
\resizebox*{10cm}{10cm}{\rotatebox{-90}{\includegraphics{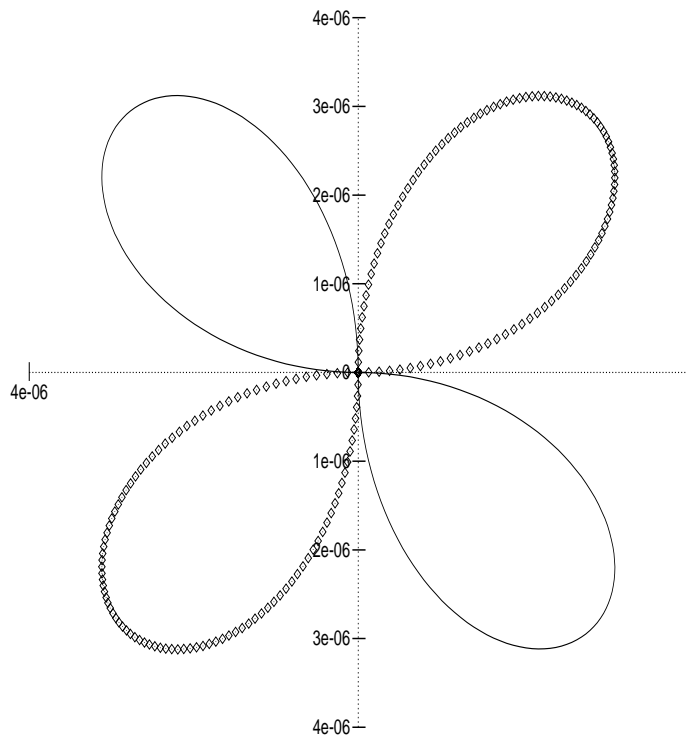}}}

\caption{Magneto scattering cross section \protect\( F_{1}(\theta )\protect \) for
a Rayleigh scatterer. The solid line is a positive correction and the points
denote a negative correction. The curve has been normalized by the parameter
\protect\( W\protect \). \label{fig2} No \textit{Hall effect} is expected
in this case because the projection onto the \protect\( y\protect \)-axis of
both corrections cancel.}
\end{figure}

\begin{figure}
\resizebox*{10cm}{10cm}{\rotatebox{-90}{\includegraphics{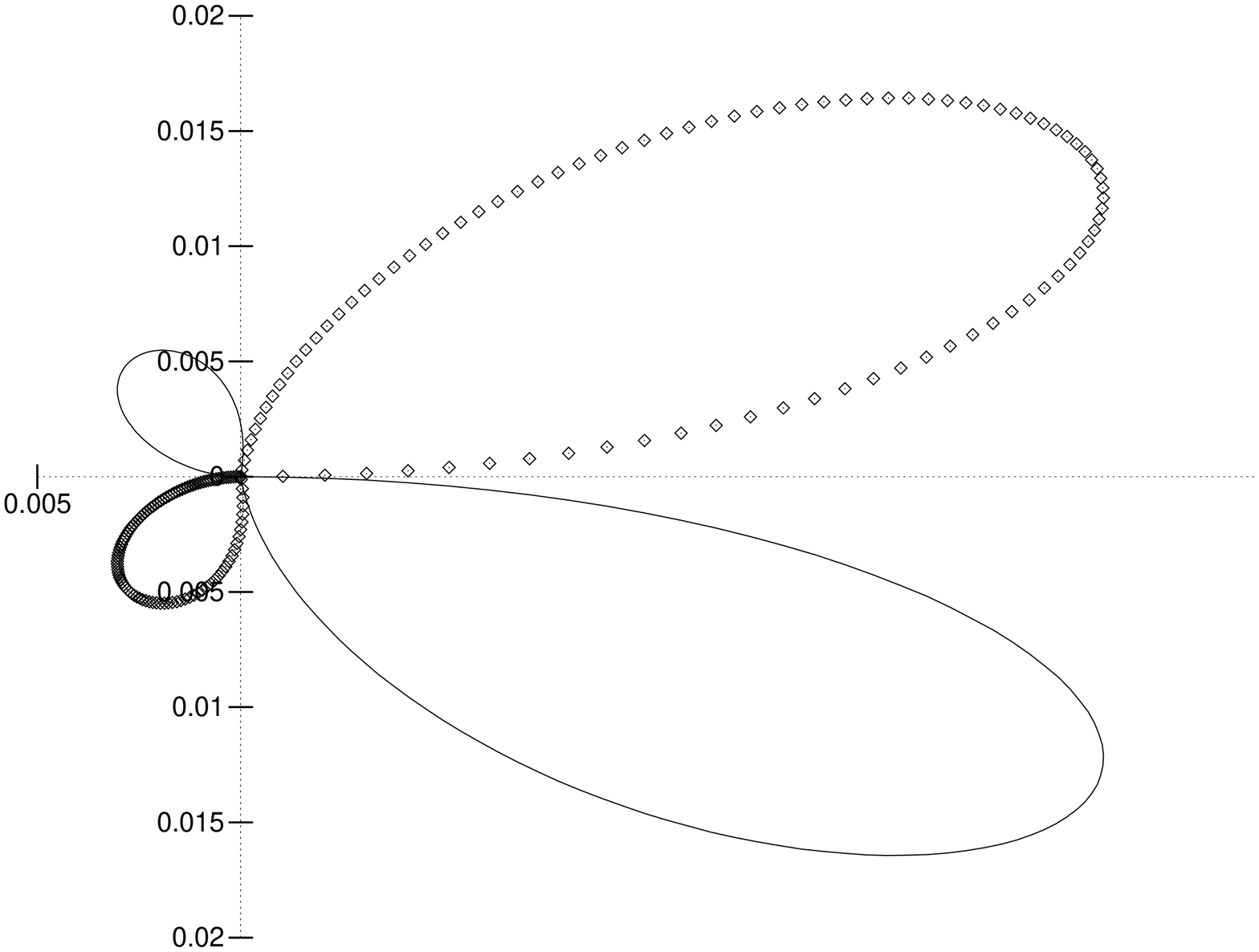}}}

\caption{Magneto scattering cross section \protect\( F_{1}(\theta )\protect \) for
a Mie scatterer of size parameter \protect\( x=2\protect \)\label{fig1}. The
curve has been normalized by the parameter \protect\( W\protect \). The solid
line is for positive correction and the points are the negative correction.
Now there is a \emph{Hall effect:} \protect\( \Delta \sigma \neq 0.\protect \)}
\end{figure}

Eq. (\ref{ttotal}) yields \( F_{1}(\cos \theta )\sim VB\cos \theta /k \) so
that Eq. (\ref{phe}) gives \( \Delta \sigma =0. \) The magneto scattering
cross section is shown in Fig. \ref{fig2} for a Rayleigh scatterer and in Fig.
\ref{fig1} for a Mie scatterer for which a non zero value of \( \Delta \sigma  \)
is seen to survive. 

\section{Stokes parameters}

To describe the flux and polarization, a 4 dimensional Stokes vector \( (I,Q,U,V) \)
can be introduced \cite{hulst}. The general relation between scattered Stokes
vector and incoming Stokes vector is,

\begin{equation}
\label{f}
(I,Q,U,V)=\mathbf{F}\, (I,Q,U,V)
\end{equation}

For a sphere and without a magnetic field, the F-matrix is well known and equals
\cite{hulst},

\begin{equation}
\label{F0}
F_{ij}^{0}=\frac{1}{k^{2}r^{2}}\left( \begin{array}{cccc}
F_{11} & F_{12} & 0 & 0\\
F_{12} & F_{11} & 0 & 0\\
0 & 0 & F_{33} & F_{34}\\
0 & 0 & -F_{34} & F_{33}
\end{array}\right) 
\end{equation}
 where 
\begin{equation}
\label{S}
\left\{ \begin{array}{c}
F_{11}=(|S_{1}|^{2}+|S_{2}|^{2})/2\, \, \, \\
F_{12}=(-|S_{1}|^{2}+|S_{2}|^{2})/2\, \, \, \\
F_{33}=(S^{*}_{2}S_{1}+S_{2}S^{*}_{1})/2\\
F_{34}=i(-S^{*}_{2}S_{1}+S_{2}S^{*}_{1})/2
\end{array}\right. 
\end{equation}
 Among these four parameters only three are independent since \( F^{2}_{11}=F^{2}_{12}+F^{2}_{33}+F^{2}_{34}. \)
The presence of the many zeros in Eq. (\ref{F0}) is a consequence of the fact
that the amplitude matrix in Eq. (\ref{amatrix}) is diagonal for one Mie scatterer.
It is in fact much more general. The form of Eq. (\ref{F0}) still holds for
an ensemble of randomly oriented particles with an internal plane of symmetry
(such as spheroids for instance) \cite{bohren}. In that case, the averaging
is essential to get the many zeros in Eq. (\ref{F0}). It also holds for a single
anisotropic particle in the Rayleigh-Gans approximation \cite{holoubek},\cite{k-filter}.

For the Mie case, the anisotropy has two consequences: the F-elements that were
zero for an isotropic particle may take finite values, and they may depend on
the azymuthal angle \( \phi  \). When a magnetic field is applied perpendicular
to the scattering plane, corrections will appear in the diagonal terms of the
amplitude matrix. We will use the vector H to denote them. When a magnetic field
is applied in the scattering plane, the amplitude matrix becomes off diagonal,
which will fill up the zeros in \( F^{0} \). We will use the vector G to denote
these new terms. 

If we call \( F^{1} \) the first-order magnetic correction to the F-matrix
one finds,

\begin{equation}
\label{F1}
F_{ij}^{1}=\frac{1}{k^{2}r^{2}}\left( \begin{array}{cccc}
H_{11} & H_{12} & \Re eG_{3} & -\Im mG_{3}\\
H_{12} & H_{11} & \Re eG_{4} & -\Im mG_{4}\\
\Re eG_{1} & \Re eG_{2} & H_{33} & H_{34}\\
\Im mG_{1} & \Im mG_{2} & -H_{34} & H_{33}
\end{array}\right) .
\end{equation}
 When \( \hat{\mathbf{B}} \) is directed along \( \hat{\mathbf{k}} \), the
G terms are given by,

\begin{equation}
\label{gbk}
G_{\hat{\mathbf{B}}=\hat{\mathbf{k}}}\, \, \, \left\{ \begin{array}{c}
G_{1}=(S^{*}_{1}R^{*}_{1}-S_{2}R_{2})/2\, \, \, \, \\
G_{2}=(-S^{*}_{1}R^{*}_{1}-S_{2}R_{2})/2\\
G_{3}=(-S^{*}_{1}R^{*}_{2}+S_{2}R_{1})/2\\
G_{4}=(S^{*}_{1}R^{*}_{2}+S_{2}R_{1})/2\, \, \, \, 
\end{array}\right. 
\end{equation}
 The general case (forward and backward directions excluded) has

\begin{equation}
\label{G}
\mathbf{G}=\frac{(\hat{\mathbf{B}}\cdot \hat{\mathbf{k}})(\hat{\mathbf{k}}\cdot \hat{\mathbf{k}}')-\hat{\mathbf{B}}\cdot \hat{\mathbf{k}}'}{(\hat{\mathbf{k}}\cdot \hat{\mathbf{k}}')^{2}-1}\mathbf{G}_{\hat{\mathbf{B}}=\hat{\mathbf{k}}'}+\frac{(\hat{\mathbf{B}}\cdot \hat{\mathbf{k}}')(\hat{\mathbf{k}}\cdot \hat{\mathbf{k}}')-\hat{\mathbf{B}}\cdot \hat{\mathbf{k}}}{(\hat{\mathbf{k}}\cdot \hat{\mathbf{k}}')^{2}-1}\mathbf{G}_{\hat{\mathbf{B}}=\hat{\mathbf{k}}}
\end{equation}
 and \( G_{\hat{\mathbf{B}}=\hat{\mathbf{k}}'} \) is obtained from \( G_{\hat{\mathbf{B}}=\hat{\mathbf{k}}} \)
by exchanging \( R_{1} \) and \( R_{2} \) in Eq. (\ref{gbk}) like in Eqs.
(\ref{tbpk},\ref{tbpkp}). Finally, we need, 
\begin{equation}
\label{H1}
\mathbf{H}=(\hat{\mathbf{B}}\cdot \hat{\mathbf{g}})\mathbf{H}_{\hat{\mathbf{B}}=\hat{\mathbf{g}}},
\end{equation}
 with 
\begin{equation}
\label{H}
\mathbf{H}_{\hat{\mathbf{B}}=\hat{\mathbf{g}}}\, \, \, \left\{ \begin{array}{c}
H_{11}=-\Im m(S_{1}Q_{1}+S_{2}Q_{2})/2\, \, \, \\
H_{12}=-\Im m(-S_{1}Q_{1}+S_{2}Q_{2})/2\\
H_{33}=\Im m(-S_{1}Q_{2}-S_{2}Q_{1})/2\, \, \, \, \\
H_{34}=\Re e(S_{1}Q_{2}-S_{2}Q_{1})/2\, \, \, \, \, \, \, \, 
\end{array}\right. 
\end{equation}

The F-matrix defined in Eq. (\ref{f}) can contain at most 7 independent constants,
resulting from the 8 constants in the amplitude matrix minus an irrelevant phase.
Our \( F^{1} \)-matrix has 12 coefficients (4 for the H vector and 8 for the
G vector). Therefore 5 relations must exist between these 12 coefficients. These
relations have not been explicitely derived.

We can write all the expressions above in a very compact way using the basis
of the Pauli matrices \cite{k-law}

\begin{equation}
\label{pauli1}
\begin{array}{c}
F_{ij}^{0}=\frac{1}{2}Tr(A^{0\dagger }\, \sigma _{i}\, A^{0}\, \sigma _{j})\\
F_{ij}^{1}=\frac{1}{2}Tr(A^{1\dagger }\, \sigma _{i}\, A^{0}\, \sigma _{j})+\frac{1}{2}Tr(A^{0\dagger }\, \sigma _{i}\, A^{1}\, \sigma _{j}),
\end{array}
\end{equation}
where \( Tr \) is the trace of the matrix, the superscript \( \dagger  \)
denotes Hermite conjugation, \( \sigma _{i} \) are Pauli matrices and \( A^{0} \),
\( A^{1} \) are zeroth and first order correction in the amplitude matrix defined
as the T-matrix from Eq. (\ref{amatrix}).

\begin{figure}
\resizebox*{10cm}{10cm}{\rotatebox{-90}{\includegraphics{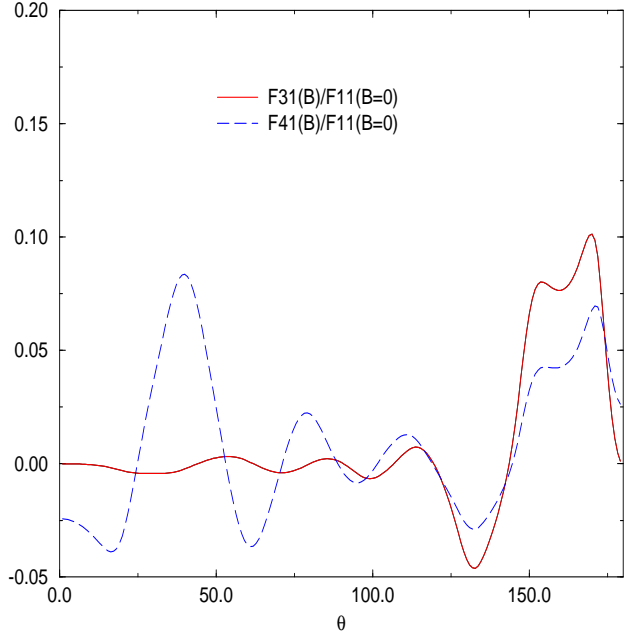}}} \resizebox*{10cm}{10cm}{\rotatebox{-90}{\includegraphics{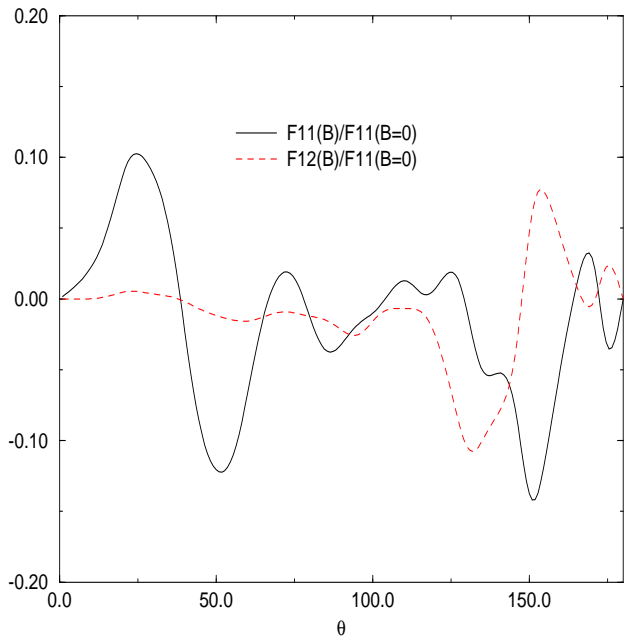}}}

\caption{Scattering matrix elements \protect\( F^{1}_{31}(B)/F^{1}_{11}(B=0)\protect \)
and \protect\( F^{1}_{41}(B)/F^{1}_{11}(B=0)\protect \) of an ensemble of water
droplets as a function of scattering angle for \protect\( \hat{\mathbf{B}}\protect \)
directed along \protect\( \hat{\mathbf{k}}\protect \) (top), and \protect\( F^{1}_{11}(B)/F^{1}_{11}(B=0)\protect \)
and \protect\( F^{1}_{12}(B)/F^{1}_{11}(B=0)\protect \) for \protect\( \hat{\mathbf{B}}\protect \)
directed along \protect\( \hat{\mathbf{g}}\protect \) (bottom). The refractive
index is 1.332, a lognormal size distribution has been used with \protect\( r_{eff}=0.75\mu m\protect \)
and \protect\( \sigma _{eff}=0.45\protect \) and \protect\( \lambda =632.8nm\protect \).
The curve has been displayed for \protect\( W=0.1.\protect \)\label{stokes}}
\end{figure}

If the incident light is unpolarized, the Stokes vector for the scattered light
is simply equal to the first column of the F-matrix in Eq. (\ref{F1}). For
instance when \( \hat{\mathbf{B}} \) is directed along \( \hat{\mathbf{k}} \),
the magnetic field will only affect \( U=F^{1}_{31} \) and the circular polarization
\( V=F^{1}_{41} \) which would be zero when no magnetic field is applied. We
choose to normalize the matrix elements \( F^{1}_{ij} \) that quantify the
deviation of the polarization from the isotropic case by the flux \( F^{0}_{11} \)
without magnetic field. In Fig. \ref{stokes} we plotted these normalized matrix
elements for the cases where \( \hat{\mathbf{B}} \) is directed along \( \hat{\mathbf{k}} \)
and where \( \hat{\mathbf{B}} \) is directed along \( \hat{\mathbf{g}} \).
We observe that off-diagonal F-elements such as \( F^{1}_{12} \) and \( F^{1}_{41} \),
are generally more important in the angle region of \( 140-170^{\circ } \),
and increase with the size parameter. In this region, these Stokes parameters
seem to be very sensitive to anisotropy as also found from studies of Stokes
parameters of quartz particles \cite{k-law}.

The F-matrix of spherical scatterers in Eq. (\ref{F0}) contains 8 zeros among
its 16 elements. This property persists for an ensemble of randomly oriented
non-spherical particles having a plane of symmetry because of the averaging
over all the orientations. In a magnetic field even spherical scatterers can
have a non zero value for these 8 elements. Furthermore we have good reasons
to believe that our theory made for spheres in a magnetic theory should also
apply to an ensemble of randomly oriented non-spherical particles in a magnetic
field, since the magnetic field direction is the same for all the particles.

We have chosen the size distribution \cite{travis} and optical parameters of
a reported experiment \cite{hovenier}, when no magnetic field is present ,
but in which all the matrix elements of \( F^{0} \) were measured and found
to be in good agreement with the theoretical evaluation from Eq. (\ref{F0}).
For water, the parameter \( W\approx 2.4\cdot 10^{-6} \) for a magnetic field
of 1T. From Fig. \ref{stokes}, we can therefore expect a modification of the
order of \( 2.4\cdot 10^{-6} \) in the region near backward scattering for
\( F^{1}_{31}(B)/F^{1}_{11}(B=0) \) when \( \hat{\mathbf{B}} \) is directed
along \( \hat{\mathbf{k}} \). The magneto-optical effects on polarization are
very small. Nevertheless they may become significant in multiple scattering,
which usually tends to depolarize completely the light.

\subsection{Forward and backward directions}

When no magnetic field is present, the situations for \( \theta =0 \) or \( \theta =\pi  \)
are similar because the scattering plane is undefined in both cases. We also
have \( H=0 \) by Eq. (\ref{H1}). The remaining contribution is therefore
only determined by the G-vector, and the final result reads for \( \theta =0 \),

\begin{equation}
\label{gt0}
F_{\theta =0}=\frac{\hat{\mathbf{B}}\cdot \hat{\mathbf{k}}}{k^{2}r^{2}}\, \left( \begin{array}{cccc}
0 & 0 & 0 & -\Im m(z)\\
0 & 0 & \Re e(z) & 0\\
0 & -\Re e(z) & 0 & 0\\
-\Im m(z) & 0 & 0 & 0
\end{array}\right) ,
\end{equation}
 with \( z=S_{1}(1)R_{1}(1). \) For \( \theta =\pi , \) 
\begin{equation}
\label{gtpi}
F_{\theta =\pi }=\frac{\hat{\mathbf{B}}\cdot \hat{\mathbf{k}}}{k^{2}r^{2}}\, \left( \begin{array}{cccc}
0 & 0 & 0 & \Im m(z')\\
0 & 0 & -\Re e(z') & 0\\
0 & -\Re e(z') & 0 & 0\\
-\Im m(z') & 0 & 0 & 0
\end{array}\right) ,
\end{equation}
 with \( z'=S_{1}(-1)R_{1}(-1). \)

The functions \( R_{1}(1) \) and \( R_{1}(-1) \) defined in Eqs. (\ref{p1},\ref{p2})
are very similar to \( S_{1}(1) \) and \( S_{1}(-1) \). Both F-matrices contain
only two real-valued independent parameters as do the corresponding T matrices.
For unpolarized incident light only the Stokes parameter \( V=F^{1}_{41}(B) \)
of these matrices is non-zero. In Fig. \ref{stokes}, all the curves are zero
for \( \theta =0 \) and \( \theta =\pi  \) except the one of \( F^{1}_{41}(B)/F^{1}_{11}(B=0) \).
In other words, unpolarized incident light will produce partially circularly
polarized light (the degree of circular polarization being precisely \( F^{1}_{41}(B)/F^{0}_{11}(B=0) \))
for \( \hat{\mathbf{B}} \) directed along \( \hat{\mathbf{k}} \) in the forward
and backward directions. This can be understood from the fact that the effective
index that one can define from Eq. (\ref{tkk}) suffers from magneto-dichroism
(\emph{i.e} different absorption for different circular polarization).

The modified reciprocity relation in the presence of a magnetic field was expressed
for the amplitude matrix in Eq. (\ref{reciprocity}). For the F-matrix it implies
exactly the different signs in the matrix elements of Eq. (\ref{gtpi}) with
respect to Eq. (\ref{gt0}).

\section{Summary and Outlook}

We have shown that the theory developed for magneto-active Mie scatterers so
far is consistent with former results concerning predictions of the light scattering
by Rayleigh scatterers in a magnetic field. Our perturbative theory provides
quantitative predictions concerning the \textit{Photonic Hall Effect} for one
single Mie sphere, such as the scattering cross section, the dependence on the
size parameter or on the index of refraction. 

Using the magneto-correction to the T-matrix we have derived the Stokes parameters
for the light scattered from a single sphere in a magnetic field. We have distinguished
two main cases. Either the magnetic field is perpendicular to the scattering
plane and there will be corrections to the usual non-zero Stokes parameters,
or when the magnetic field is in the scattering plane, the corrections fill
up the F-matrix elements which were previously zero. We have discussed the particular
cases of forward and backward scattering.

We hope that these results will be useful in comparing them to the situation
in multiple scattering. Even after many scattering events, we suspect that the
presence of a magnetic field prevents the Stokes parameters \( U,V \) and \( Q \)
to be zero. In single scattering, their order of magnitude is controlled by
the parameter \( W=V_{0}B\lambda  \). In multiple scattering however, this
parameter must be replaced by \( fV_{0}Bl^{*} \), where \( f \) is the volume
fraction of the scatterers and \( l^{*} \) the transport mean free path, and
\( fl^{*}\gg 1 \). We expect to find more significant effects in this case.

We thank Geert Rikken and Joop Hovenier for useful comments. We thank the referees
for their work, and in particular for mentioning the work of Kuz'min \emph{etal}.

\appendix 

\section{Appendix: Derivation of reciprocity and parity relations}

In the indices for polarization in the T-matrix, the state of helicity \( \sigma  \)
is to be referred to the direction of the wave vector immediately close to it.
In Eq. (\ref{parity},\ref{reciprocity}), \( T_{-\mathbf{k}\sigma ,-\mathbf{k}'\sigma '} \)
for instance really means \( T_{-\mathbf{k}\sigma (-\mathbf{k}),-\mathbf{k}'\sigma '(-\mathbf{k}')}. \)
To derive these equations, we start from Eq. (\ref{T1}) in which we change
both incoming and outcoming wave vectors into their opposite:

\begin{equation}
\label{T2}
\mathbf{T}_{-\mathbf{k}\sigma ,-\mathbf{k}'\sigma '}^{1}=\sum _{J,M,\lambda }\, (-M)\, \left[ \alpha _{J,\lambda }\, \mathbf{Y}_{J,M}^{\lambda }(-\hat{\mathbf{k}})\cdot \chi _{\sigma (-\mathbf{k})}(-\mathbf{k})\, \mathbf{Y}^{\lambda *}_{J,M}(-\hat{\mathbf{k}}')\cdot \chi _{\sigma '(-\mathbf{k}')}(-\mathbf{k}')\right] ,
\end{equation}
 where \( \alpha _{J,\lambda } \) is a well known coefficient, \( \chi _{\sigma (\mathbf{k})} \)
is the eigenfunction of the operator \( \mathbf{S}\cdot \mathbf{k} \) with
eigenvalue \( \sigma (\mathbf{k}) \) the helicity. \( \mathbf{S} \) is a spin
one operator acting on three-dimensional vectors. The summation is to be performed
for \( \lambda =e,m \) only, which are associated with the two transverse components
of the given vector spherical harmonics. \( \mathbf{Y}^{\lambda }_{JM}(\mathbf{k}) \)
are well-defined linear combinaisons of \( \mathbf{Y}_{J,J}^{M}(\mathbf{k}) \),\( \mathbf{Y}_{J,J-1}^{M}(\mathbf{k}) \)
and \( \mathbf{Y}_{J,J+1}^{M}(\mathbf{k}) \) that obeys \cite{newton}

\[
\mathbf{Y}_{JM}^{\lambda }(\mathbf{k})\cdot \mathbf{k}=0,\, \, \, \, \, \, \lambda =e,m.\]

We now use the relations,
\begin{equation}
\label{YM}
\begin{array}{c}
\mathbf{Y}_{J,M}^{e}(-\hat{\mathbf{k}})=(-1)^{J+1}\mathbf{Y}_{J,M}^{e}(\hat{\mathbf{k}})\\
\mathbf{Y}_{J,M}^{m}(-\hat{\mathbf{k}})=(-1)^{J}\mathbf{Y}_{J,M}^{m}(\hat{\mathbf{k}})
\end{array}.
\end{equation}
 The eigenfunctions \( \chi _{\sigma (\mathbf{k})} \) also change under parity
since 
\[
\chi _{\sigma (-\mathbf{k})}(-\mathbf{k})=-\chi _{\sigma (\mathbf{k})}(\mathbf{k}).\]
 Because of this additional minus sign, the parities of the vector spherical
harmonics are in fact,

\begin{equation}
\label{YMP}
\begin{array}{c}
\mathbf{PY}_{J,M}^{e}=(-1)^{J}\mathbf{Y}_{J,M}^{e}\\
\, \, \, \, \mathbf{PY}_{J,M}^{m}=(-1)^{J+1}\mathbf{Y}_{J,M}^{m}
\end{array}.
\end{equation}
 The parity symmetry relation of Eq. (\ref{parity}) follows from the application
of these relations into Eq. (\ref{T2}). The proof of the reciprocity symmetry
relation of Eq. (\ref{reciprocity}) is similar, where now the following relations
are necessary

\begin{equation}
\label{conj}
\mathbf{Y}_{J,M}^{\lambda *}=(-1)^{J+M}\mathbf{Y}_{J,-M}^{\lambda }
\end{equation}
 for \( \lambda =e,o,m \) and 
\[
\chi ^{*}_{\sigma (\mathbf{k})}(\mathbf{k})=\chi _{-\sigma (\mathbf{k})}(\mathbf{k}).\]
 The change of sign of \( \mathbf{B} \) is provided by the \( M \) factor
in Eq. (\ref{T2}) as surmised.

\end{document}